\documentclass[twocolumn,letter,11pt]{article}
\usepackage{amsmath,bm}
\usepackage{caption}
\usepackage{siunitx}
\usepackage{enumitem}
\usepackage{array}
\usepackage{graphicx}
\usepackage{amssymb}
\usepackage[margin=1in]{geometry}
\usepackage{cite}
\usepackage{booktabs}
\usepackage{multirow}
\usepackage{authblk}  
\usepackage{url}
\usepackage{titlesec}
\usepackage[dvipsnames,table,xcdraw]{xcolor}
\usepackage[colorlinks=true, allcolors=blue]{hyperref}

\title{Acoustic-Assisted Fabrication of Thin Shells with Spatially Distributed Imperfections}

\author[1]{Ilyes Krida}
\affil[1]{Department of Mechanical and Aerospace Engineering, University of Houston}
\author[2]{Jacob Tang}
\affil[2]{Department of Aerospace \& Mechanical Engineering, University of Southern California}
\author[1]{Daniel Floryan}
\author[1*]{Tian Chen}
\affil[*]{Corresponding author: tianchen@uh.edu}

\date{}                

\begin{document}
\maketitle

\textit{
Thin-shell structures, found in biological systems such as beetle carapaces and widely used in aerospace, civil, and mechanical engineering, achieve remarkable strength-to-mass ratio given their slenderness and curved geometries. However, their load-bearing capacity is highly sensitive to geometric imperfections, which are often unavoidable during fabrication and can trigger subcritical buckling. Silicone-based hemispherical domes have served as an experimental modal system to study this phenomenon, yet prior work has largely focused on localized dimples or flat imperfections, failing to capture the spatially distributed nature of real-world imperfection patterns.
Here, we introduce an acoustic-assisted method for fabricating thin shells with spatially distributed, vibrational mode-shaped imperfections. Silicone is cast onto a thick elastic mold excited by a speaker, and vibration-induced flow during curing creates thickness variations. High-speed imaging and $\mu$CT scanning reveal accumulation of material at the antinodes of the mold’s vibrational modes. We show that imperfection geometry can be tuned by acoustic frequency, while their amplitude increases with acoustic volume. Buckling experiments demonstrate significant reductions in critical pressure, offering a scalable platform to study and tune imperfection-sensitive mechanics. Beyond shell mechanics, we offer a scalable and tunable fabrication method for patterning soft materials in applications ranging from morphable surfaces to bioinspired design.
}

\section*{Introduction}

Thin-shell structures are ubiquitous in both biological and engineered systems with examples span a wide range of length scales including virus capsids, pollen grains, beetle exoskeletons, pressure vessels, spacecrafts and civil infrastructure \cite{lidmar2003virus, katifori2010foldable, adriaenssens2014shell, pellegrino2015folding}. The mechanical advantage of thin shells stems from their curved geometry and large radius-to-thickness ratios. However, this slenderness also makes some shells highly susceptible to \textit{subcritical buckling}, a sudden and catastrophic instability often triggered by small geometric imperfections.

This imperfection sensitivity has posed a fundamental challenge in shell mechanics for over a century. Classical elastic stability theory, from Zoelly's analytical solution~\cite{zoelly1915ueber} to Koiter's post-buckling analysis \cite{koiter1967stability}, predicts the critical buckling loads for shells with perfect geometries. Yet, experimentally measured buckling pressures of fabricated shells are often significantly lower-sometimes by over 80\%. This has been shown to be predominately due to small geometric imperfections unavoidable during manufacturing processes \cite{godoy1996thin, hutchinson1967imperfection}. To reconcile theory with experiments, a knockdown factor defined as the ratio between the observed buckling pressure and the theoretical prediction for a perfect shell, has become a key empirical metric~\cite{von1939buckling, nasto2014localized}. Yet, it remains an active research topic on what ``physically realistic'' geometric imperfections most severely reduce the knockdown factor~\cite{derveni2025most}.

The recent resurgence of interest in shell buckling is partly driven by the advent of novel fabrication and experimental techniques that enable the controlled introduction of geometric imperfections into otherwise near-perfect shells~\cite{hutchinson2016buckling}. Rather than experimenting with full-scale structures, desktop-sized silicone casts have been adopted as an experimental model system~\cite{lee2016fabrication,lee2016geometric}. These experimental studies have focused on introducing localized imperfections such as dimples or thickness reductions, fabricated using indentation or flexible molds \cite{jimenez2017technical, marthelot2017buckling,yan2020buckling,abbasi2021probing, abbasi2023comparing}. These model systems have demonstrated how imperfection amplitude, location, and symmetry influence buckling behavior, and have been well-supported by Finite Element (FE) simulations and reduced-order shell theories \cite{lee2016geometric, jimenez2017technical}. While recent works have begun investigating multi-imperfection behavior~\cite{baizhikova2024uncovering,derveni2025defect}, it is difficult for such fabrication methods to capture the complex, global imperfection fields that arise in physical shells due to manufacturing variability. It is also not practical to construct dedicated fabrication setups for all possible imperfection geometries.

Here, we introduce an acoustic-assisted fabrication approach to create thin hemispherical shells with prescribed global imperfection fields shaped by the vibrational modes of their casting mold. 
Vibrational excitation has long been used to impose pattern in physical systems. Chladni plate experiments, where fine particles accumulate along the nodal lines of a vibrating surface, illustrate how modal patterns emerge from standing wave excitation~\cite{rossing2013principles}. In fluid systems, Faraday waves are standing waves on a fluid surface excited by vertical vibration that create rich patterns through nonlinear resonances and symmetry breaking~\cite{douady1988pattern}. In manufacturing, vibration has been harnessed to assist in powder spreading, colloidal self-assembly, and material redistribution in soft or particulate media \cite{wu2020vibration, baudoin2020acoustic,kopitca2021programmable,derayatifar2024holographic,ledda2025fluid}. Central to these effects is the coupling between oscillatory driving and the medium's response, often mediated by steady secondary streaming flows that persist over longer timescales than the driving frequency \cite{perinet2017streaming}.

\begin{figure}[ht!]
\centering
\includegraphics [width=80mm]{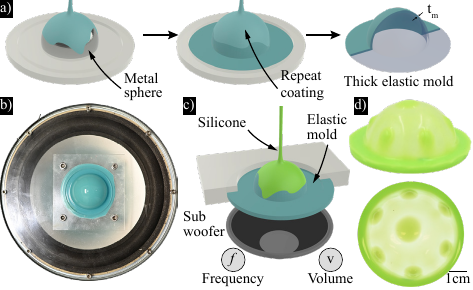}
\caption{\textbf{Overview of the casting and acoustic-fabrication protocol.} a) Fabrication of an elastic mold by repeated coating of a metal sphere. b) Photograph of the mold attached to a PMMA plate bolted to the speaker. c) Casting of a silicone hemispherical shell atop the elastic mold as it is being excited acoustically at a known frequency, $f$ and volume, $v$. d) An example of a resulting imperfect shell with volume maximized to make visible the imperfections.}
\label{fig:1}
\end{figure}

In our setup, by actively exciting a thick elastic hemispherical mold using a speaker with tunable frequencies and volume, we induce resonant modal deformation of the mold surface during the curing of a thin liquid silicone layer. This vibration generates spatially structured flows within the liquid silicone, causing material to accumulate at antinodes and to thin elsewhere, thereby imprinting the mold's vibrational displacement field as a permanent variation in shell thickness.

We characterize the resulting imperfection fields using $\mu$CT scanning and stereographic optical imaging, reconstructing the full scalar field of shell thickness as a function of spherical coordinates. We find that these fields qualitatively resemble the vibrational eigenmodes as predicted by FE simulations, with modal symmetry and spatial distribution closely preserved. While the geometry of the imperfections are mode-dependent, the amplitude of thickness variation can be tuned continuously by varying acoustic excitation volume. We then perform mechanical buckling experiments under vacuum loading to measure the effect of these globally distributed imperfections on structural stability.

\section*{Results}

To fabricate thin hemispherical shells with spatially distributed imperfections, we begin by casting a hemispherical elastic mold through repeated coating of a precisely machined metal sphere ($D=\SI{50.8}{\milli\metre}$) with silicone (Mold Star 16 Fast Platinum Silicone Rubber) (Fig.~\ref{fig:1}a)~\cite{lee2016fabrication}. By allowing the silicone to cure in-between successive coat, the mold thickness \( t_\mathrm{m} \) increases per coat by approximately $\SI{0.35}{\milli\metre}$. Once we achieve a mold thickness that exceeds the thickness of the eventual hemispherical shells by an order of magnitude, we proceed with casting. 
The hemispherical elastic mold is rigidly mounted onto a circular PMMA plate ($\approx \SI{230}{\milli\metre}$ in diameter) with a circular cutout in the center matching the mold's diameter. This plate is bolted to the rim of an acoustic speaker (Polk Audio PSW10 10'' Powered Subwoofer) to form an enclosed volume (Fig.~\ref{fig:1}b). The frequency and volume of speaker is digitally controlled with a PC. See SI Sec 2.1 for detailed fabrication protocol of the elastic mold.

\begin{figure}[ht!]
\centering
\includegraphics[width=80mm]{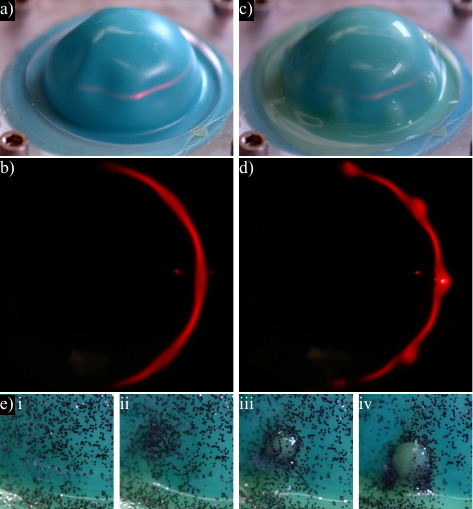}
\caption{
\textbf{Physics of silicone casting under acoustic excitation.}
a) Shell excited at a high speaker volume to make visible the periodic vibration as captured by a slow motion camera (see SI Video 1). b) Using a laser sheet and photographing the vibrating mold from top down, we capture the locations of the nodes and antinodes at a single elevation. c) Once silicone is poured onto the vibrating mold, multiple bumps are formed (see SI Video 2). d) It is observed that the bumps are formed at the antinodes of the mold using laser sheet photography. e) Stroboscopic slow motion capture (frame rate equally to the frequency of the speaker) shows a sequence of fluid flow tracked using reflective particles, i–iv captures progressive accumulation of material at antinodes (See SI Video 3).
}
\label{fig:2}
\end{figure}

To allow easy separation after casting, a mold-release (Ease Release 200, Smooth-on) is first sprayed on the elastic mold. Following this, a two-part Vinyl Polysiloxane (Elite Double 32, Zhermack) is mixed and poured onto the elastic mold to sufficiently cover the entire mold. The speaker is then powered on at a specific frequency and volume (Fig.~\ref{fig:1}c). The acoustics induces vibration in the mold which redistributes the still-liquid silicone into a new steady-state prior to solidification (See SI Video 1). 
As the silicone cures, this redistribution becomes permanently imprinted in the shell geometry. An example of such spatially distributed imperfections is shown in Fig.~\ref{fig:1}d. By tuning the frequency and the volume, we are able to fabricate shells of different imperfection geometries.
Note that the timescale of curing ($\sim \SI{20}{\minute}$) is much slower than the formation of the imperfection pattern ($\sim \SI{3}{\second}$), which is itself much slower than the period of the excitation ($\le \SI{0.01}{\second}$). 
Further, the exponential nature of curing suggests the silicone remains predominately liquid in the initial time-period after pouring~\cite{lukin2020platinum}. See SI Sec 2.2 for detailed fabrication protocol of the elastic mold.

\begin{figure*}[ht!]
\centering
\includegraphics[width=165mm]{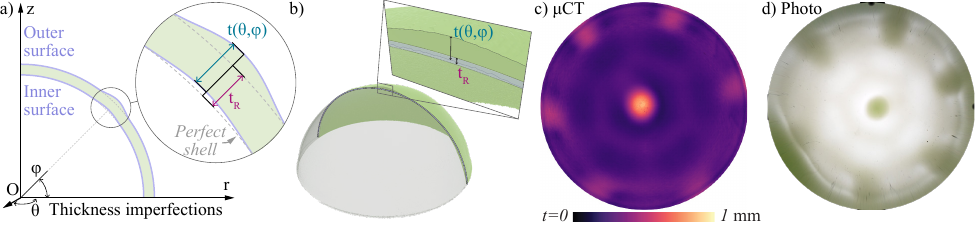}
\caption{\textbf{Quantification of acoustically generated imperfection as a measure of shell thickness.} a) Schematic of thickness measurement per spherical coordinate, b) Reconstructed 3D geometry of an imperfect shell based on $\mu$CT scanning, one section of a bump-like imperfection is shown, c) thickness profile as mapped to a disk showing deviation of thickness from a perfect shell, d) photography based capture and mapping of thickness deviation.}
\label{fig:3}
\end{figure*}

We proceed to understand the formation of imperfections. Specifically, we use slow motion videography to capture how the vibration of the mold shapes the liquid silicone that is poured over (See SI Video 2). We choose one modal frequency \( f = 167\,\text{Hz} \) as informed by FE analysis featuring 8 nodes and 8 antinodes distributed around the circumference. Adjusting the speaker volume, we observe periodic deformations on the mold matching the modal shape from the FE (Fig.~\ref{fig:2}a). By setting the frequency of video capture to match that of the speaker frequency, the resulting stroboscopic video shows negligible movement, demonstrating that the shell is vibrating in sync with the speaker (See SI Video 2). A laser sheet is projected across the shell of radius $r$ at a given elevation, $z=0.25r$, to highlights the location of nodes and antinodes by recording the distortions along the intersection between the elastic mold and the laser (Fig.~\ref{fig:2}b).

Once silicone is poured onto the mold, the onset of acoustic-driven vibration causes the uncured silicone to flow and accumulate at distinct points (Fig.~\ref{fig:2}c). The same laser sheet photography from the top view shows that the accumulations coincide with the locations of the antinodes (Fig.~\ref{fig:2}d). This seems counterintuitive as the antinodes experience the largest vibration amplitude and could cause the liquid to flow away. To further understand the dynamics of this flow, we introduce reflective particles on the surface of the uncured silicone. In addition to an oscillatory primary flow whose frequency is equal to that of the mold vibration, slow-motion imaging captures the progressive migration of particles toward antinodes during curing from all directions on the surface (Fig.~\ref{fig:2}e), driven by a secondary mean (streaming) flow. Once the bumps are formed over a time period $\sim \SI{5}{\second}$ that is much longer than the period of the vibration, an internal toroidal circulation is observed from the motion of the particles. Similar to vortex rings, the liquid silicone moves to the circumference of the bump before moving under and emerging at the apex (See SI Video 3). This secondary streaming flow is much weaker than the primary oscillatory flow, and bears resemblance to the streaming flow associated with Faraday waves~\cite{perinet2017streaming}.
If the speaker is shut off or tuned to another frequency, the fluid migrates accordingly until curing occurs in approximately \SI{20}{\minute}, at which point the shell is peeled off the mold. This allows us to correlate the modal vibration to the eventual shape of the shells cast upon the elastic mold.

To quantify the geometry of the imperfect shells fabricated using the above method, we reconstruct their 3D shape using micro-computed tomography ($\mu$CT) and optical photography. We observe the acoustically generated imperfection generally takes the form of thickness variations across the surface of the shell (Fig.~\ref{fig:3}a). Further, deviations largely occur on the outer surface of the shell, \textit{e.g.}, the inner surface remains largely spherical. We quantify the thickness of an imperfect shell at every spherical coordinate as $t(\theta, \phi) = d_\mathrm{outer}(\theta, \phi)-d_\mathrm{inner}(\theta, \phi)$ where $d_\mathrm{outer}$ and $d_\mathrm{inner}$ are the distances from the outer and inner surfaces of the shell to the origin respectively. See SI Sec 3.1 for detailed $\mu$CT parameters.

By processing the image stack (Fiji: ImageJ), \(\mu\)CT scanning provides an unambiguous 3D model of both the inner and outer surfaces as meshes. By sampling these two meshes at fixed $\theta$ and $\phi$ coordinates, we are able to reconstruct the thickness variation across the shell, with $t_\mathrm{max}=\SI{0.77}{\milli\metre}$ and $t_\mathrm{min}=\SI{0.056}{\milli\metre}$ in this example where the speaker volume is maximized to make visible the imperfections (Fig.~\ref{fig:3}b). We proceed to map $t(\theta, \phi)$ onto a 2D disk with evenly spaced angular and radial polar coordinates (Fig.~\ref{fig:3}c). The result exhibits modal symmetry in general. In this particular imperfect shape, six localized bumps are observed close to the equator of the shell, and one bump is present at the pole. In the intermediate areas, several regions of thinning are measured in a hexagonal pattern. 

As \(\mu\)CT is not suitable for structures at all length scales, we simplify the process using optical photography as a measuring tool. We position each fabricated shell atop a LED light panel with constant ambient lighting. A digital camera (Nikon D780, 105mm f/2.8G) is placed vertically above the shell with a separation distance of \SI{400}{\milli\metre}. Each resulting image is processed to remove optical distortion and then stereographically mapped to 2D (Fig.~\ref{fig:3}d)~\cite{lisle2004stereographic}. The degree of light penetration through the silicone serves as an indirect measure of thickness. The exposure is set to ensure no highlight or shadow clipping, \textit{i.e.}, no pure white or black, then we are able to fit the relationship between intensity and thickness with an attenuation coefficient~\cite{baker1984effect}. Therefore, if we provide independent measures of the maximum and minimum thicknesses ($t_\mathrm{max}, t_\mathrm{min}$), we can infer the entire thickness field (See SI Sec 3.2 and 3.3 the image processing algorithm).
These findings confirm that the acoustic excitation protocol introduces repeatable, spatially structured imperfections whose geometry closely resembles vibrational mode shapes. See SI 2.3 for a library of different imperfection patterns.

\begin{figure}[ht!]
\centering
\includegraphics[width=80mm]{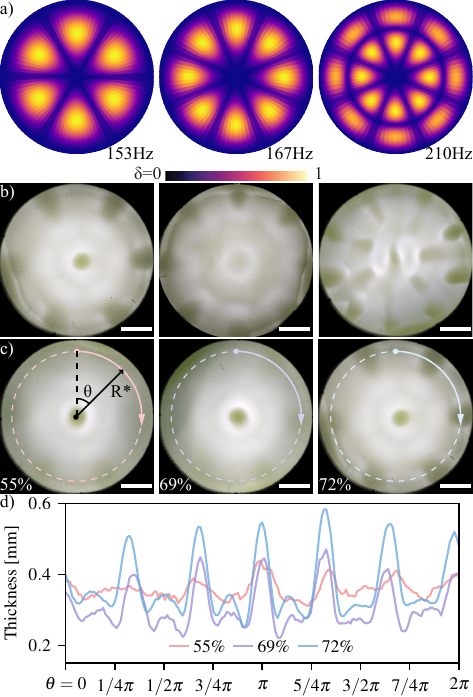}
\caption{
Comparison between Finite Element (FE) predicted vibrational modes and the experimentally realized imperfection patterns in acoustically excited hemispherical shells at the modal frequencies.
a) Simulated modal shapes for three distinct vibrational modes, showing radial deviation fields mapped into 2D plane. b) Top and side views of hemispherical shells showing global shape deviations and mode symmetry. c) d) By varying the speaker volume and setting the frequency constant at \SI{153}{\hertz}, we can tune the amplitude of the thickness field $t(\theta,\phi)$ with changing its characteristics. The stereographic photograph of each shell is shown. d) The thickness distribution at one elevation, $\phi=\arcsin{1/6}, \theta=[0,2\pi]$ is plotted for shells fabricated with volumes at 55\%, 69\% and 72\% of maximum. All scale bars are \SI{1}{\centi\metre}.}
\label{fig:4}
\end{figure}

We have shown earlier that the elastic mold vibrates in sync with the frequency of the speaker (Fig.~\ref{fig:2}a). This allows us to use FE simulations to predict and engineer the vibrational modes of the elastic mold as a function of the speaker frequency. Specifically, we calculate the natural vibrational modes and frequencies of the elastic mold with the equatorial degrees-of-freedom constrained. We select three different modal shapes for experimentation, namely $f=153$, $167$ and $\SI{210}{\hertz}$. The first two exhibit a single ring of six and eight antinodes respectively. The last modal shape exhibits two rings of 10 antinodes. See SI 4 for detailed FE analysis protocol.

\begin{figure*}[ht!]
\centering
\includegraphics[width=165mm]{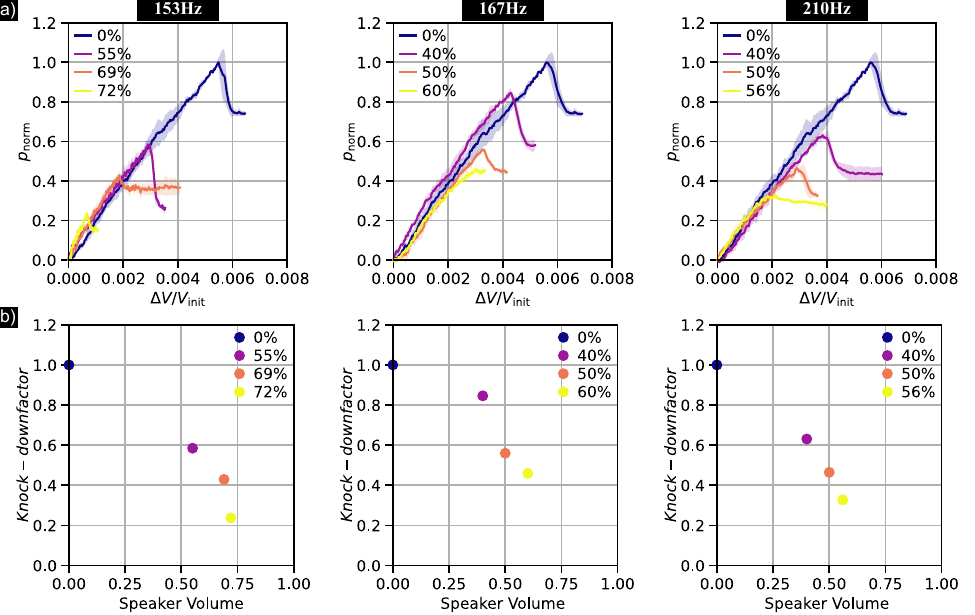}
\caption{\textbf{The buckling response of of hemispherical shells with acoustic-imperfections.} a) Normalized pressure-volume curves for shells fabricated with specific speaker excitation volume and frequencies. Behavior of a shell with no acoustic imperfections is plotted for reference.
b) Knock-down factors as a function of speaker volume, quantifying the reduction in structural stability with increasing imperfection severity.}
\label{fig:5}
\end{figure*}

For each mode, the nodal displacements from the undeformed configuration are normalized as the absolute value of the radial displacement with the maximum deviation equaling to 1 (Fig.~\ref{fig:4}a). We project the resulting displacement fields to 2D for visual comparison. 
We then set the speaker to the corresponding frequencies and cast the shells using the aforementioned method. The resulting fabricated shells (Fig.~\ref{fig:4}b) exhibit mode-symmetric bump-like deformations whose quantity and spatial orientation closely mirror the corresponding simulated vibrational modes. 
Note the visual comparison is aimed at correlating the displacement field of a vibration mold to the thickness variation of a shell cast atop that mold. These observations support that acoustic excitation can reliably embed modal features as geometric imperfections in thin-shell. The differences between the photographed specimens and the modal shapes is largely due to the fact that the speaker volume exceeds significantly from the linearized regime. Full-field non-linear dynamical analysis will be conducted in the future to quantitatively construct this correlation.

Next, we begin to modulate the speaker volume, $v=[55\%,69\%,72\%]$, while keeping the frequency constant at $f=\SI{153}{\hertz}$. The resulting shells exhibit the same imperfection patterns qualitatively, however, the amplitude of thickness variation increases systematically (Fig.~\ref{fig:4}c). Thickness measurement from $\mu$CT scanning confirms this behavior. Plotting the thickness distribution along a single latitude $\phi=\arcsin{1/4}$ shows six peaks in all three fabricated shells. However, their amplitude increases systematically as a function fo the speaker volume  (Fig.~\ref{fig:4}d). By modulating both frequency and volume, we are able to systematically change imperfection geometry and severity.

Lastly, we quantify the mechanical response of this class of imperfect shells under increasing vacuum pressure (Fig.~\ref{fig:5}).
In addition to perfect shells, a total of nine different imperfect hemispherical shells are fabricated and divided into three groups, each subjected to acoustic excitation at a fixed frequency listed in Fig.~\ref{fig:4}a. Within each group, the shells are exposed to varying excitation amplitudes by adjusting the speaker's output volume. Quasi-static under pressure experiments are performed on the shells. The interior of each shell is connected to an automated syringe pump via a flexible tube. The volume in the shell $V_\mathrm{shell}$, the tube $V_\mathrm{tube}$ and the \SI{1}{\milli\litre} syringe $V_\mathrm{syr}$ forms a closed system with a total volume $V_\mathrm{total}$. A pressure sensor is used to record pressure as the syringe draws air and expands the total volume (See SI Sec 5.1 for detailed experimental setup). As the total volume increases, the pressure decreases until the shell buckles, at which point the volume decreases abruptly and the pressure rebounds.

Assuming ideal gas law, we can isolate the volume under the shell (See SI Sec 5.2 for calculation). This allows us to define a normalized volume $V_\mathrm{norm}$ as $\Delta V / V_{\mathrm{shell}}$ where $\Delta V= V_\mathrm{shell}-{V_\mathrm{shell}}^\mathrm{init}$. 
We further define a normalized internal pressure as $p_{\mathrm{norm}}=p/p_\mathrm{crit}$ where $p_\mathrm{crit}$ is the under pressure required to buckle a perfect hemispherical shell. We proceed to plot the normalized internal pressure as a function of normalized volume change (Fig.~\ref{fig:5}a). Each $PV$ curve corresponds to a specific excitation volume and frequency. Increasing the speaker volume results in a systematic reduction in critical buckling pressure and earlier onset of instability. The stiffness of all shells remain relatively constant.

We then calculate the corresponding knock-down factors of each shell as the ratio between the buckling pressure of imperfect shells and that of the perfect one (See SI Sec 5.3 for characterization of a perfect shell).
These are plotted against speaker volume (Fig~\ref{fig:5}b). The plots confirm a consistent decline in structural capacity with stronger perturbation volume.

\section*{Discussion}
We have introduced an acoustic-assisted fabrication method to create hemispherical shells with spatially distributed, mode-shaped geometric imperfections in the form of thickness variations. By casting liquid silicone atop a vibrating elastic mold, we exploit vibration-driven flow and secondary streaming to imprint modal shapes directly onto the shell's thickness field. The resulting imperfections are tunable in geometry via excitation frequency and scalable in severity through speaker volume. Quantitative measurements from $\mu$CT and optical imaging confirm the high fidelity of modal patterning, while mechanical buckling tests demonstrate a systematic reduction in structural capacity with increasing imperfection amplitude. This technique opens new opportunities for studying multi-mode and nonlinear imperfection interactions, as well as programmable buckling responses. One may extend this method to fabricate morphable surfaces, acoustically controlled soft robots, or artifical shells with programmed buckling modes. More broadly, the integration of vibration with soft matter fabrication offers a versatile route for patterning complex geometries in thin-film and elastomeric systems.

\bibliographystyle{unsrt}  
\bibliography{references}  

\end{document}